\UseRawInputEncoding
\documentclass[
  reprint,
  superscriptaddress,
  nofootinbib,
  nobibnotes,
  amsmath,amssymb,
  aps,
  prb
]{revtex4-2}
\usepackage{float}
\usepackage{capt-of}
\usepackage{graphicx}
\usepackage{epstopdf}
\usepackage{changepage}
\usepackage{dcolumn}
\usepackage{bm}
\usepackage{gensymb}
\usepackage{upgreek}
\usepackage{hyperref}
\usepackage{tabularx}
\usepackage{xcolor}
\usepackage{xparse,xcoffins}
\usepackage{verbatim}
\usepackage{xurl}
\usepackage[export]{adjustbox}

\preprint{APS/123-QED}

\begin{document}

\title{Single crystal growth, structural and magnetic properties of CeZn$_{2-x}$Ga$_{2+x}$}

\author{Danila Sokratov}
\affiliation{Maryland Quantum Materials Center, Department of Physics, University of Maryland, College Park, MD 20742, USA}

\author{H. Cein Mandujano}
\affiliation{Department of Chemistry and Biochemistry, University of Maryland, College Park, MD 20742, USA}
\affiliation{Maryland Quantum Materials Center, Department of Physics, University of Maryland, College Park, MD 20742, USA}

\author{Ram Kumar}
\author{Jared Z. Dans}
\author{Phineas Sobel}
\author{Nicholas A. Crombie}
\affiliation{Maryland Quantum Materials Center, Department of Physics, University of Maryland, College Park, MD 20742, USA}

\author{Alicia Manj\'on-Sanz}
\affiliation{Neutron Scattering Division, Oak Ridge National Laboratory, Oak Ridge, TN 37831, USA}

\author{Danielle R. Yahne}
\affiliation{Neutron Scattering Division, Oak Ridge National Laboratory, Oak Ridge, TN 37831, USA}

\author{Philip Piccoli}
\affiliation{Department of Geological, Environmental, and Planetary Sciences, University of Maryland, College Park, MD 20742, USA}

\author{Peter Y. Zavalij}
\affiliation{Department of Chemistry and Biochemistry, University of Maryland, College Park, MD 20742, USA}

\author{Efrain E. Rodriguez}
\affiliation{Department of Chemistry and Biochemistry, University of Maryland, College Park, MD 20742, USA}
\affiliation{Maryland Quantum Materials Center, Department of Physics, University of Maryland, College Park, MD 20742, USA}

\author{Johnpierre Paglione}
\email{paglione@umd.edu}
\affiliation{Maryland Quantum Materials Center, Department of Physics, University of Maryland, College Park, MD 20742, USA}
\affiliation{Canadian Institute for Advanced Research, Toronto, Ontario M5G 1Z8, Canada}

\date{\today}

\begin{abstract}

The tetragonal BaAl$_4$ ($I4/mmm$) parent structure underpins a diverse family of materials exhibiting novel phenomena, including nematic superconductivity, topological semimetallicity, and heavy fermion behavior. The recent growth of ternary R-Zn-Ga compounds, such as the previously reported CeZn$_2$Ga$_2$, has explored some of the members exhibiting rare-earth magnetism within this family. In this paper, we report on the structural and magnetic properties of single crystals of CeZn$_{2-x}$Ga$_{2+x}$, a Ga-rich analogue of CeZn$_2$Ga$_2$. Our CeZn$_{2-x}$Ga$_{2+x}$ samples exhibit magnetic properties distinct from the paramagnetic behavior previously reported for CeZn$_2$Ga$_2$. We observe a magnetic transition around 4 K, pronounced metamagnetic states at low temperatures, and strong magnetic anisotropy. Though there are batch-to-batch variations that suggest a strong sensitivity to local structural imperfections, we consistently see the presence of magnetic transitions and metamagnetic states in our crystals. To investigate the local structural sensitivity hypothesis, we performed Reverse Monte Carlo analysis of collected powder neutron diffraction data, revealing the presence of significant local crystallographic disorder of the magnetic Ce site. Our findings demonstrate that the positional disorder drives competing ferromagnetic and antiferromagnetic correlations that lead to the observed spin glass behavior and complex anisotropic magnetism. This study illustrates that tuning the local crystallographic disorder enables engineering frustrated magnetic states in BaAl$_4$-type and similar intermetallic structures.

\end{abstract}

\maketitle

\section{Introduction}

\begin{figure}[!ht]
    \adjincludegraphics[height=11.2cm,trim={1.5cm 1cm 0.5cm 1cm},clip]{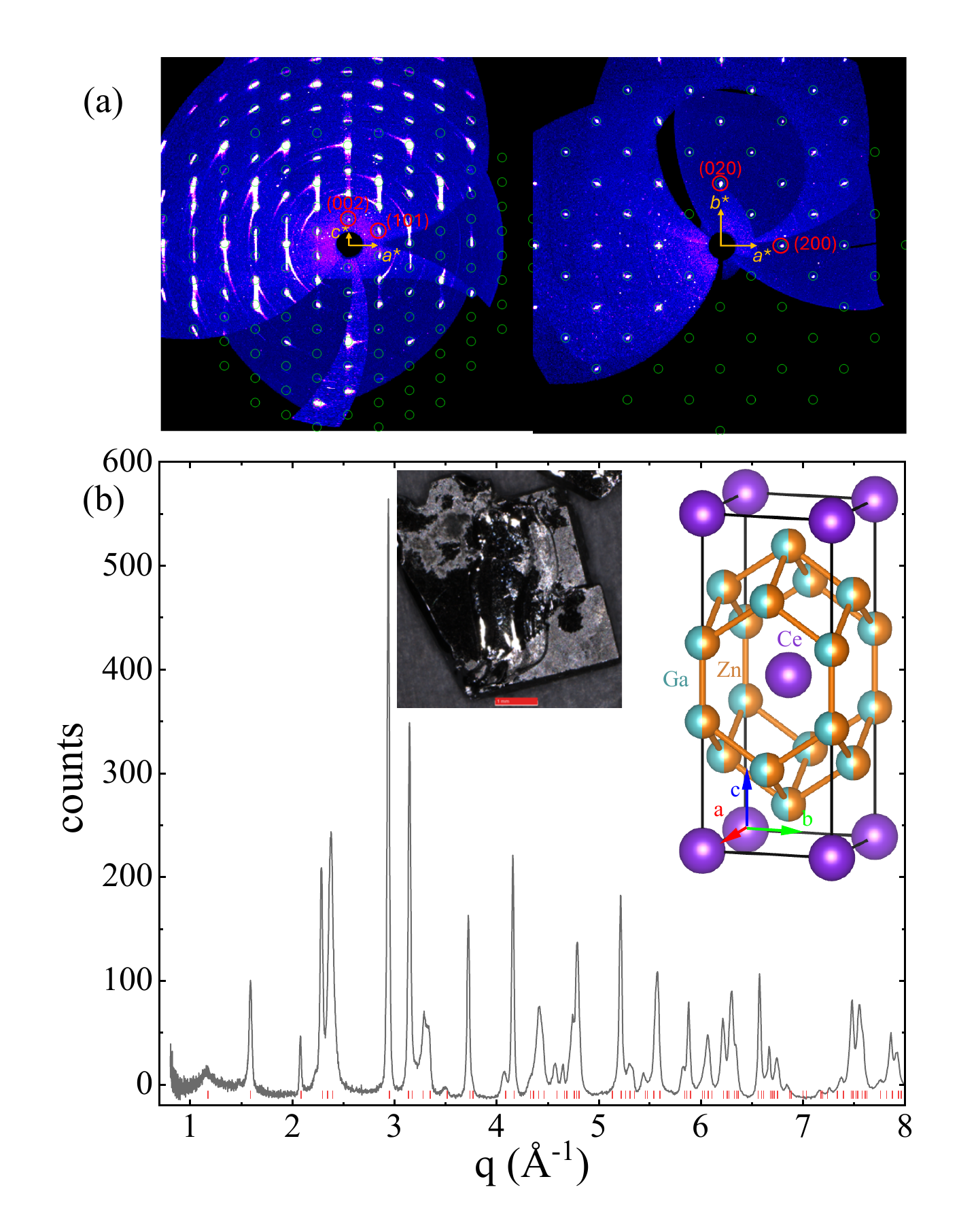}
    \captionof{figure}{(a) Precession images from single crystal X-ray scattering for \emph{hk0} (top, right) and \emph{h0l} (top, left). (b) Neutron powder diffraction data obtained at POWGEN. Background subtracted data shown in gray with peak indices obtained from ideal CeZn$_2$Ga$_2$ structure reported in ref.\cite{verbovytskyy_novel_2010}. Typical crystal shown in inset, together with a unit cell of 1-2-2 structure.}
    \label{fig:powgen_sxrd}
\end{figure}

The spin glass phase has been a topic of interest since the first experimental signatures observed in AC-susceptibility measurements of AuFe alloys\cite{cannella_magnetic_1972}. There, a sharp cusp-like feature was detected, indicative of a magnetic phase transition. The $T_c$ of this transition depended on the concentration of Fe atoms that act as magnetic impurities in the structure. Over the following 50 or so years, many models have been proposed and iterated over to explain this state, including multivalley sketches\cite{mezard_sg_1986}, short-range spin droplet spin glass models\cite{fisher_ordered_1986} and spin-density wave pictures\cite{cable_neutron-polarization-analysis_1982, cable_neutron-polarization-analysis_1984}. Despite a prolonged period of research, the spin glass state is still not fully understood and continues to attract scientific attention, including topics such as window and ferroic glasses\cite{sherrington_bzt_2013}, as well as social\cite{kirkpatrick_social_2012} and neural networks\cite{bouchaud_application_2023, andriushchenko_new_2022}.

The BaAl$_4$ parent structure (space group $I4/mmm$) is a fertile ground for discovering novel quantum phenomena, notably hosting materials exhibiting exotic states such as unconventional superconductivity (e.g., the prototypical heavy-fermion superconductors CeCu$_2$Si$_2$\cite{Steglich1979, Rauchschwalbe1982}, or nematic phases in certain iron pnictides\cite{paglione_high-temperature_2010, Chu2010, Stewart2011}), robust topological states\cite{wang_crystalline_2021}, charge-density waves\cite{Saraf2025}, and complex magnetic ground states\cite{Aoki1993}. Within this family, R-Zn-Ga systems have been recently explored, with previous studies reporting the synthesis and characterization of RZn$_2$Ga$_2$ single crystals\cite{verbovytskyy_novel_2010}. Specifically, CeZn$_2$Ga$_2$ was reported to be paramagnetic down to 1.7 K, despite exhibiting high-temperature ferromagnetic fluctuations noted from Curie-Weiss analysis. In these 1-2-2 Ce-based compounds, the ground state has been shown to be determined by the position on the Doniach phase diagram\cite{DONIACH1977231}, defined by the competition of the Ruderman-Kittel-Kasuya-Yosida (RKKY) interaction and the Kondo effect. The ground state depends on the exchange coupling constant $J$, and so depending on the $f$-electron hybridization strength, the systems manifest as magnetically tunable paramagnets\cite{raymond_magnetic_1999}, quantum critical antiferromagnets that can be driven to superconductivity with applied pressure\cite{mathur_magnetically_1998}, or heavy-fermion superconductors\cite{Steglich1979}. The presence of ferromagnetic fluctuations while the system stays paramagnetic down to 1.7K suggests that this CeZn$_2$Ga$_2$ system exists near a magnetic instability or Quantum Critical Point (QCP).

Motivated by the potential for complex and sensitive magnetism, we aimed to grow single crystals of CeZn$_2$Ga$_2$ using a self-flux method. In this work, we find that CeZn$_{2-x}$Ga$_{2+x}$ single crystals exhibit magnetic properties starkly different from the previously reported paramagnetic CeZn$_2$Ga$_2$. We observe a magnetic transition around 4 K, intricate metamagnetic transitions at low temperatures, and significant magnetic anisotropy between $H\parallel c$ and $H\perp c$. Furthermore, these magnetic features, including the transition temperature, display notable batch-to-batch variations, strongly hinting at the influence of local structural deviations. Similar sensitivity to site disorder and crystalline imperfections has been observed to drive spin glass behaviors in isostructural compounds like PrAu$_2$Si$_2$\cite{Li_2022}, and to tune complex metamagnetic behavior in related Ce intermetallics like Au-deficient CeAuBi$_2$\cite{Hodovanets2026} and CeCoIn$_5$\cite{PaglioneCeCoIn5}. We hypothesize that these complex magnetic phenomena, particularly the observed spin glass-like behavior, arise not primarily from the Zn/Ga site disorder, but from small displacements of the Ce atom from its ideal central position within the unit cell.

Here, we present a comprehensive investigation of single-crystalline CeZn$_{2-x}$Ga$_{2+x}$, combining detailed magnetic susceptibility, heat capacity, and resistivity measurements with structural insights gained from X-ray and neutron diffraction data. Through reverse Monte Carlo analysis, we illuminate the nature of local Ce displacements and their direct correlation with the emergent spin glass behavior and anisotropic metamagnetic states, highlighting the critical role of subtle positional disorder in dictating complex magnetic phases in these intermetallic compounds.

\section{Methods}

\subsection{Single crystal growth}

Single crystal samples of nominally CeZn$_2$Ga$_2$ were grown with self-flux in a similar manner to previous reports\cite{verbovytskyy_novel_2010}. In order to investigate effects of rare earth purity on crystal quality, stoichiometry and size, two different sources of Ce were used across batches of growths. High purity Ce (Ames lab, Alfa Aesar rod 99.9$\%$ metals basis), Zn (Alfa Aesar shot 99.999$\%$ metals basis) and Ga (thermoscientific pellets 99.9999$\%$ metals basis) were combined in a ratio of $1:2:10$ with excess Ga acting as a flux at high temperatures. The mixture was placed inside of an alumina Canfield crucible and then sealed inside a quartz tube in an argon atmosphere. The resulting ampules were heated to $800\degree C$ at max furnace rate and held for an hour, then further heated to $900\degree C$ and held for 24 hours, before cooling down at a rate of $5\degree \frac{C}{hr}$ to $450\degree C$, where the ampules were centrifuged to get rid of excess Ga flux. This growth technique results in large plate-like single crystals, typically $2\times 2\times 0.5 mm^3$ with short axis being along the $c$-axis. The grown crystals are often limited in size by the crucible used. Additional details of growth conditions, elemental sources and their purity and their effects on the crystal growths are outlined in the Supplemental Information.

The chemical composition of the crystals was verified by using a combination of energy-dispersive X-ray spectroscopy (EDS), wavelength-dispersive spectroscopy (WDS) and X-ray fluorescence spectroscopy (XRF) techniques. EDS and EDS analyses were performed using a JEOL JXA-8900R Superprobe at the University of Maryland. EDS was used to perform qualitative analyses, and compositions were subsequently verified using WDS, where x-ray counts were corrected using a ZAF algorithm\cite{Armstrong1991}. XRF measurements were performed in an iXRF Atlas M spectrometer equipped with a Rh source and Si3N4 detectors.

Detailed elemental analyses by EDS, WDS, and XRF revealed that our single crystals possess a chemical composition between CeZnGa$_2$ to CeZnGa$_3$, which varies between technique and also crystal growth batch. Complicating the chemical composition issues, powder X-ray and neutron diffraction experiments struggle refining the exact composition and crystal structure, though the $I4/mmm$ BaAl$_4$-type structure reported in ref.\cite{verbovytskyy_novel_2010} indexes the peak positions nearly perfectly. Our best structural refinements of the neutron powder diffraction data gives a chemical concentration of Ce$_{0.9}$Zn$_{1.6}$Ga$_{2.4}$, though the exact occupancies are uncertain as the best refinement is only at $\approx$12 wr\%. This fits with the Zn deficiency observed with elemental analysis techniques. We note that identifying the Zn deficiency difference between the extremes of CeZn$_2$Ga$_2$ and CeZnGa$_3$ is difficult as the one electron difference does not show up in measurements like X-ray diffraction or fluorescence.

\begin{figure*}[!ht]
    \adjincludegraphics[height=9.5cm,trim={2cm 1cm 0cm 1cm},clip]{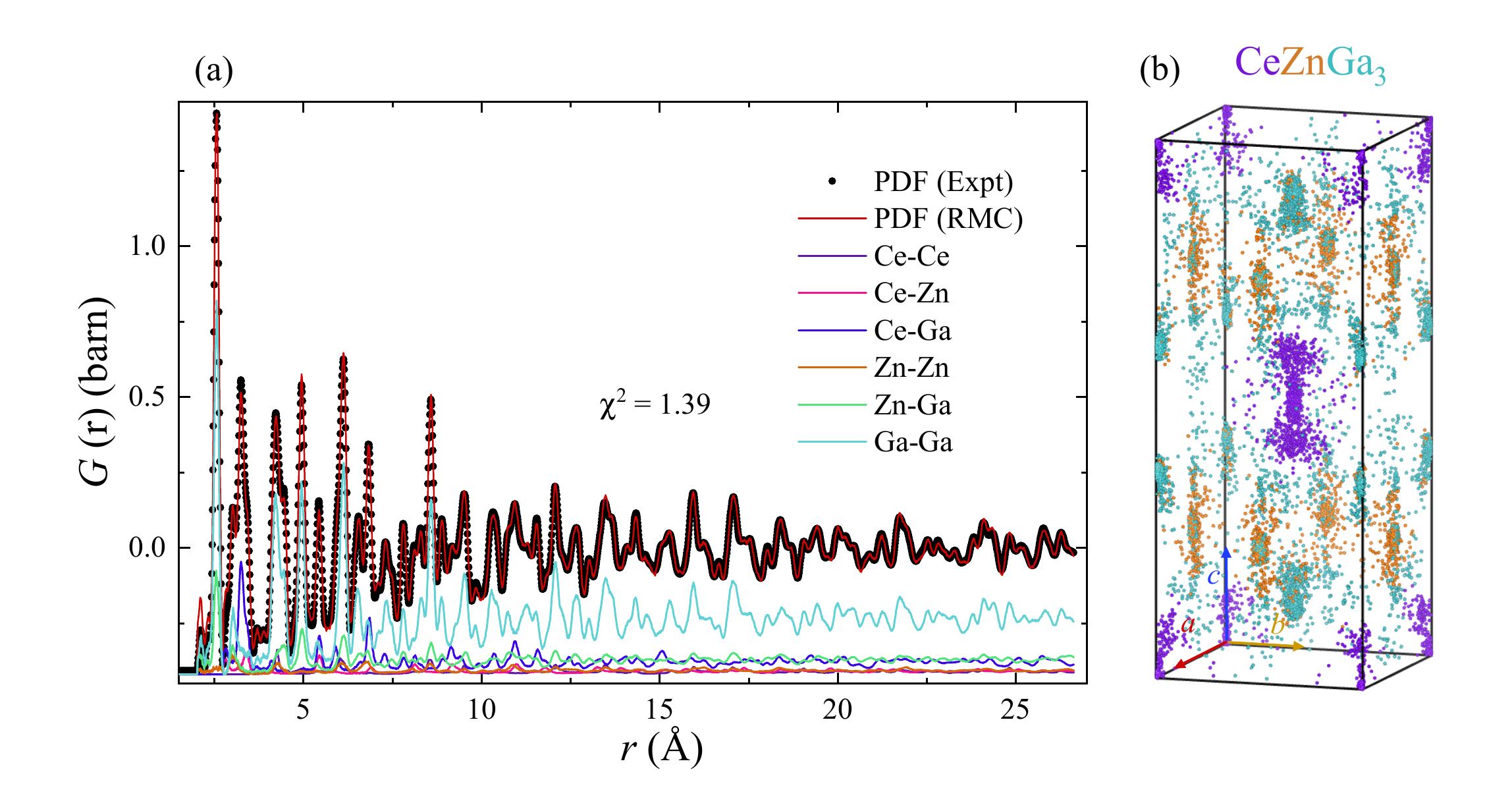}
    \captionof{figure}{(a) Reverse Monte Carlo fit to the experimental Pair Distribution Functions (PDF) obtained from POWGEN measurement. Individual PDFs from atomic bonds are shown below the fit. Calculation done assuming a fixed CeZnGa$_3$ chemical composition. (b) Average unit cell obtained by collapsing the multi-cell structure into a single cell for visual purposes. Purple dots indicate positions of Ce atoms, orange dots of Zn atoms, and teal of Ga atoms. The unit cell generally agrees with the idealized 1-2-2 structure, though there is notable spread in the Ce atomic position in the center of the unit cell.}
    \label{fig:rmc}
\end{figure*}

\subsection{Transport measurements}

Electrical transport and heat capacity measurements were made with a Quantum Design Physical Properties Measurement System (PPMS). For the transport measurements, excess Ga flux was removed by polishing, and the samples were shaped into typical rectangular bars. Electrical contacts were made with gold wires that were attached with DuPont 4929 silver paste in a 4-wire configuration. Temperature and field-dependent magnetic measurements were carried out with a Quantum Design Magnetic Properties Measurement System (MPMS). Due to the anisotropic amplitude of magnetic moment of the samples and the preferred growth orientation of the crystals, measurements with field perpendicular to the $c$-axis were typically carried out on a quartz rod with small amounts of GE varnish to minimize background, while the measurements with field parallel to the $c$-axis were typically carried out in a straw to avoid the sample torquing out of position. $\chi_{dc}(T)$ measurements were conducted in standard zero-field cooled cooling (ZFC) and field cooled (FC) modes. Consecutive $\chi_{dc}(T)$ measurements at different fixed fields and $M(H)$ at fixed temperatures were all started in the ZFC mode by first warming the sample well into the paramagnetic (PM) regime around 100 K, oscillating the field to get rid of any remanent field, and then cooling down to the measurement temperature.

\begin{figure*}[!ht]
    \adjincludegraphics[height=4.7cm,trim={3cm 1cm 0cm 1cm},clip]{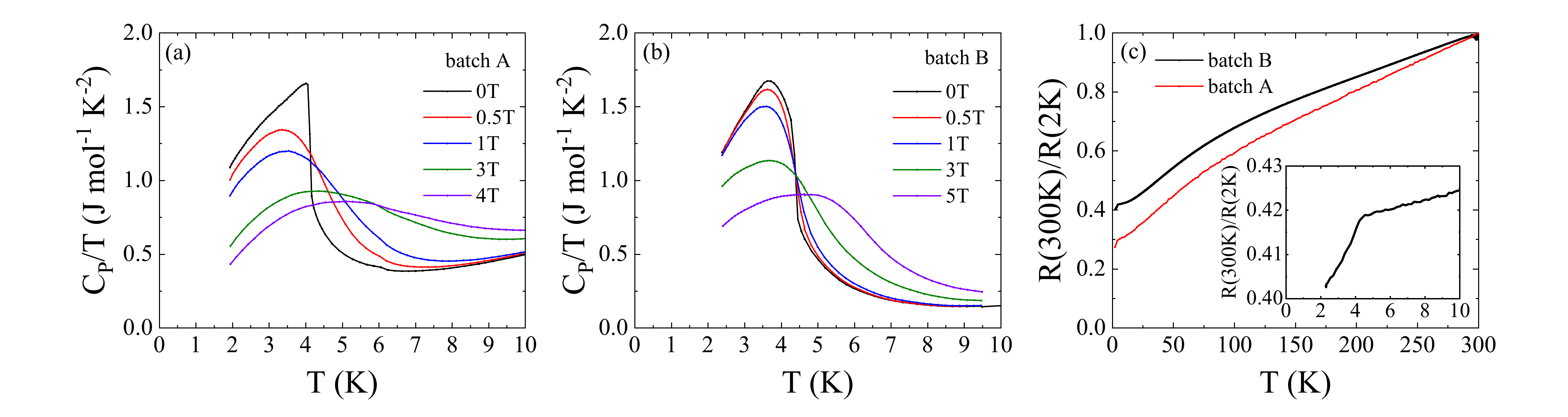}
    \captionof{figure}{Heat capacity and resistivity of CeZn$_{2-x}$Ga$_{2+x}$. Heat capacity of batch A (a) and batch B (b) at fixed fields. (c) Normalized resistance of batch A and batch B samples. Inset shows zoom lf low temperature magnetic transition in batch B sample.}
    \label{fig:hc-res}
\end{figure*}

\subsection{Diffraction measurements}

The structural analysis was done with a Rigaku MiniFlex600 X-ray diffractometer with X-ray  diffraction (XRD) patterns taken on a powder with monochromated Cu-K$\alpha$ radiation ($\lambda \sim$ 1.5406 \AA). Samples were ground into powder from as-grown single crystals in agate mortars. Powder X-ray plots are shown in SI. Small single crystals were measured using a D8 Venture with Photon III detector, shown in Fig.\ref{fig:powgen_sxrd}(a). Effects of annealing on crystal properties and X-ray diffraction patterns were explored and are discussed in SI.

Neutron powder diffraction measurements were carried out on various batches of grown single crystals. Measurements on three different batches were performed on the HB-2A diffractometer at Oak Ridge National Laboratory, and on a different batch for the POWGEN beamline. No batches were mixed due to slight differences in diffraction peaks from powder X-ray measurements. Neutron powder diffraction (NPD) measurements on the HB-2A (POWDER) diffractometer at Oak Ridge National Laboratory were performed using an incident beam wavelength of 1.54 \AA (Ge115 monochromator) with collimator settings open-open-12”. Single crystals were powderized and loaded into a 6mm aluminum can, backfilled with 1 atm He gas, and loaded into an orange cryostat with base temperature of ~1.5 K. Full diffraction patterns were taken at 1.5, 5, 50, and 300 K and refined using the GSASII refinement software\cite{toby_gsas-ii_2013}. Additional NPD data were collected on the POWGEN beamline at the Spallation Neutron Source at Oak Ridge National Laboratory\cite{Huq:in5025}. Approximately 2 g of powder were loaded into an 8 mm diameter vanadium can and placed in the POWGEN automatic sample changer. Data were collected at 10 and 300 K using central wavelengths of $\lambda$ = 0.8 \AA (covering a d-spacing range of 0.15-7.76 \AA) and $\lambda$ = 2.665 \AA (covering a d-spacing range of 1.11-20.44 \AA) for 1 h and 30 min each measurement.

\section{Results}

\subsection{Characterization}

\subsubsection{Neutron and X-ray diffraction}

We have performed detailed powder neutron diffraction experiments on various batches of the CeZn$_{2-x}$Ga$_{2+x}$ compound to investigate the crystal structure and potential occupancies of the Zn and Ga atoms in the structure. A subset of the POWGEN collected data with peak indices from the idealized CeZn$_2$Ga$_2$ tetragonal $I4/mmm$ structure is shown in Fig.\ref{fig:powgen_sxrd}. The peak positions are in good agreement with the expected structure, though we were unable to adequately refine the data from either the HB2A or the subsequent POWGEN measurements. The broadness of the [002] (first) peak in Fig.\ref{fig:powgen_sxrd} highlights that there is a significant strain on the center Ce atom in the unit cell. 

We have performed X-ray diffraction on both single crystals and powder samples obtained from grinding single crystals on various batches of the CeZn$_{2-x}$Ga$_{2+x}$ compound. The single crystal precession images are shown in Fig.\ref{fig:powgen_sxrd}. The diffraction patterns for powder samples (shown in SI) are qualitatively similar to those previously reported\cite{verbovytskyy_novel_2010}, though we do see significant differences in the relative amplitudes of a few major peaks. We note that these relative amplitudes differ between batches and are affected by annealing, as well as the source of Ce used for the growths. Results closest to the reported relative amplitudes\cite{verbovytskyy_novel_2010} do not involve annealing and use Ce purchased from Ames lab. The latter suggests the potential for disorder from presence of other elements as impurities in Ce, though we have been unable to detect them. We believe that the source of these amplitude changes could be resulting from the displacement of the Ce atoms from the average structure. The X-ray and neutron refinements cannot conclusively solve the average structure of these crystals, though peak indexing and a good fit of high R PDF data obtained from POWGEN is an indication that the $I4/mmm$ structure is a close approximation (shown in SI).

\subsubsection{Reverse Monte Carlo}

To improve on the diffraction analysis of the neutron data, we performed reverse Monte Carlo calculations in an attempt to better understand the formed structure, shown in Fig.\ref{fig:rmc}. The fit assumes the chemical composition CeZnGa$_3$ suggested by chemical composition measurements. The results suggest that the structure found in ref.\cite{verbovytskyy_novel_2010} of tetragonal $I4/mmm$ for the CeZn$_2$Ga$_2$ structure is still on average the correct unit cell, though importantly, we do see a splitting of the Ce site location into a dumbbell-shape, seen in Fig.\ref{fig:rmc} (b). This site displacement suggests that the local disorder is causing the structure to be glass-like. This is consistent with the observed strain on the (002) peak observed in the POWGEN data, as well as results from annealing of the crystals. If the local structure is allowed to fluctuate where the Ce atom lands, then leaving the crystal to anneal for a prolonged period of time will shift the positions of the Ce atoms, but it would still on average be displaced from the center of the ideal unit cell. Observed changes in the diffraction pattern and magnetic properties of samples before and after annealing are consistent with the above discussion. We believe that the anisotropic magnetic properties and the competition of FM and AFM orders in the structure are likely resulting from this local change in the Ce atom positions. This is further supported by the observed bifurcation in the ZFC and FC curves, which suggests a possible spin glass structure at low temperatures and fields.

\subsection{Heat capacity and resistivity}

Previous reports of the CeZn$_2$Ga$_2$ compound\cite{verbovytskyy_novel_2010} report a paramagnetic behavior down to base temperature of 1.78 K, despite the presence of ferromagnetic fluctuations indicated by $\theta \sim 4.4$ K extracted from the Curie-Weiss fit above 100 K. We have performed detailed heat capacity and resistivity measurements to study the behavior of our CeZn$_{2-x}$Ga$_{2+x}$ single crystals, shown in Fig.\ref{fig:hc-res}. We observe sensitive magnetism in this system and suspect it may have been missed with the slightly larger applied magnetic fields in previous studies. To highlight its sensitivity, we discuss two exemplary batches of the formed single crystals, grown with the same temperature profile described above but with different sources of Ce. 

\begin{figure}[!ht]
    \includegraphics[width=0.99\linewidth]{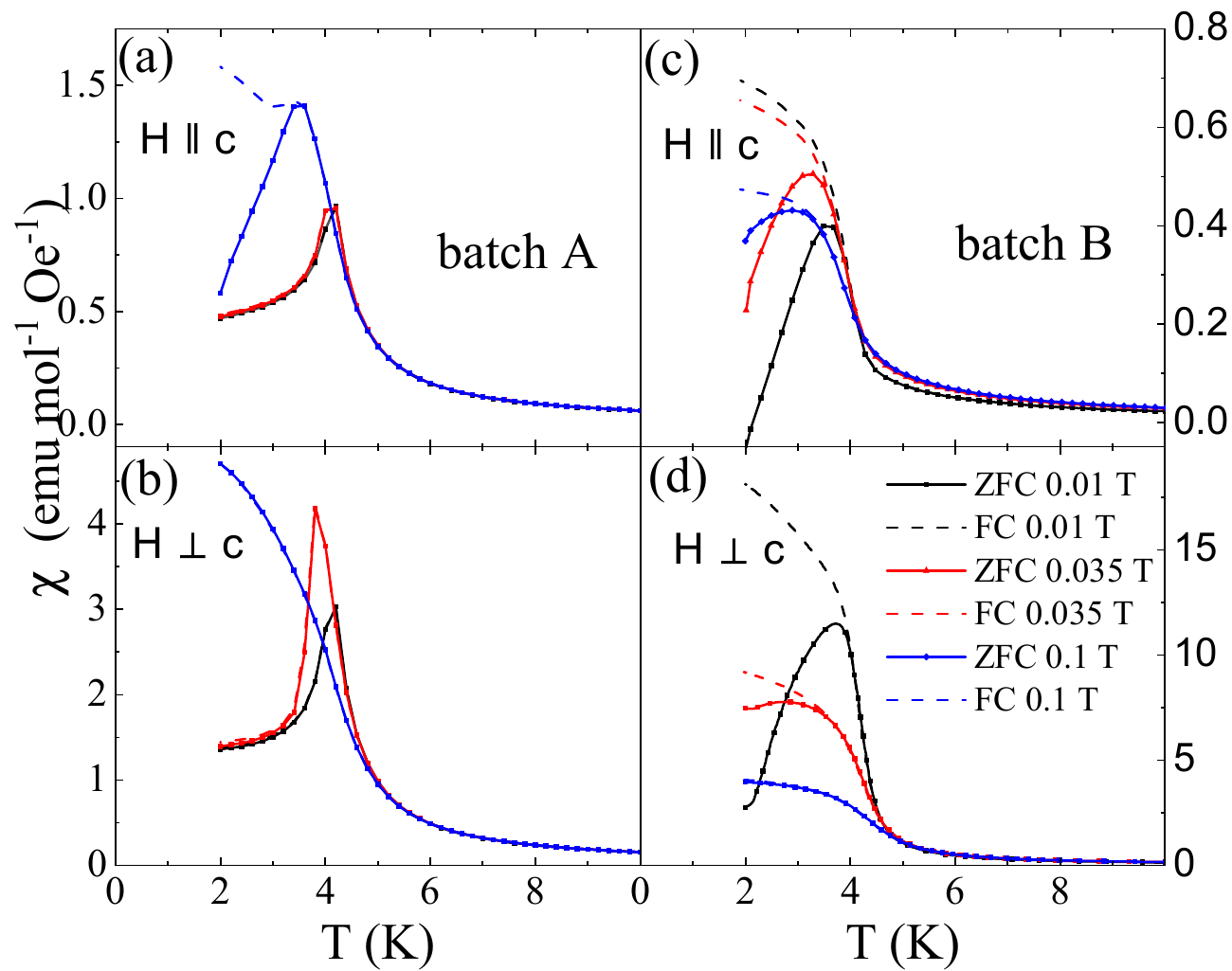}
    \captionof{figure}{Magnetic susceptibility $\chi$ of CeZn$_{2-x}$Ga$_{2+x}$ compared between two batches.  Low temperature $\chi$ for field (a) parallel to and (b) perpendicular to the $c$-axis-axis at fixed temperatures for batch A. Low temperature $\chi$ for field (c) parallel to and (d) perpendicular to the $c$-axis-axis at fixed temperatures for batch B.}
    \label{fig:chi}
\end{figure}

The heat capacity signal of batch A shows a sharp transition $\sim 4$ K at zero field, which generally broadens and its peak shifts up in temperature with increasing field. A similar transition can be observed in heat capacity signal of batch B, though it is smoother and broader at zero field, compared to batch A. Similarly, the transition broadens and its peak shifts up in temperature as a function of increasing field. The resistivity signal between the two batches is very similar, though the particular crystals show a difference in the resistivity residual ratio (RRR). For an easier comparison, the normalized resistance is shown in Fig.\ref{fig:hc-res} (c). The qualitative response between batches is nearly identical, and we observe a kink in resistance around 4 K in both batches. This transition is not found to be sensitive to magnetic fields for fields under 1~T, consistent with the heat capacity response. We note that there is no significant shift in transition temperature seen in resistivity under 1.3 GPa of pressure applied in a clamp cell setup, shown in SI. The RRR of measured samples is consistently below 5, suggesting a presence of disorder or other sources of increased scattering. This is consistent with the strain observed in the structure, as detected by neutron scattering and single crystal X-ray precession images.

\begin{figure}[!ht]
    \includegraphics[width=0.99\linewidth]{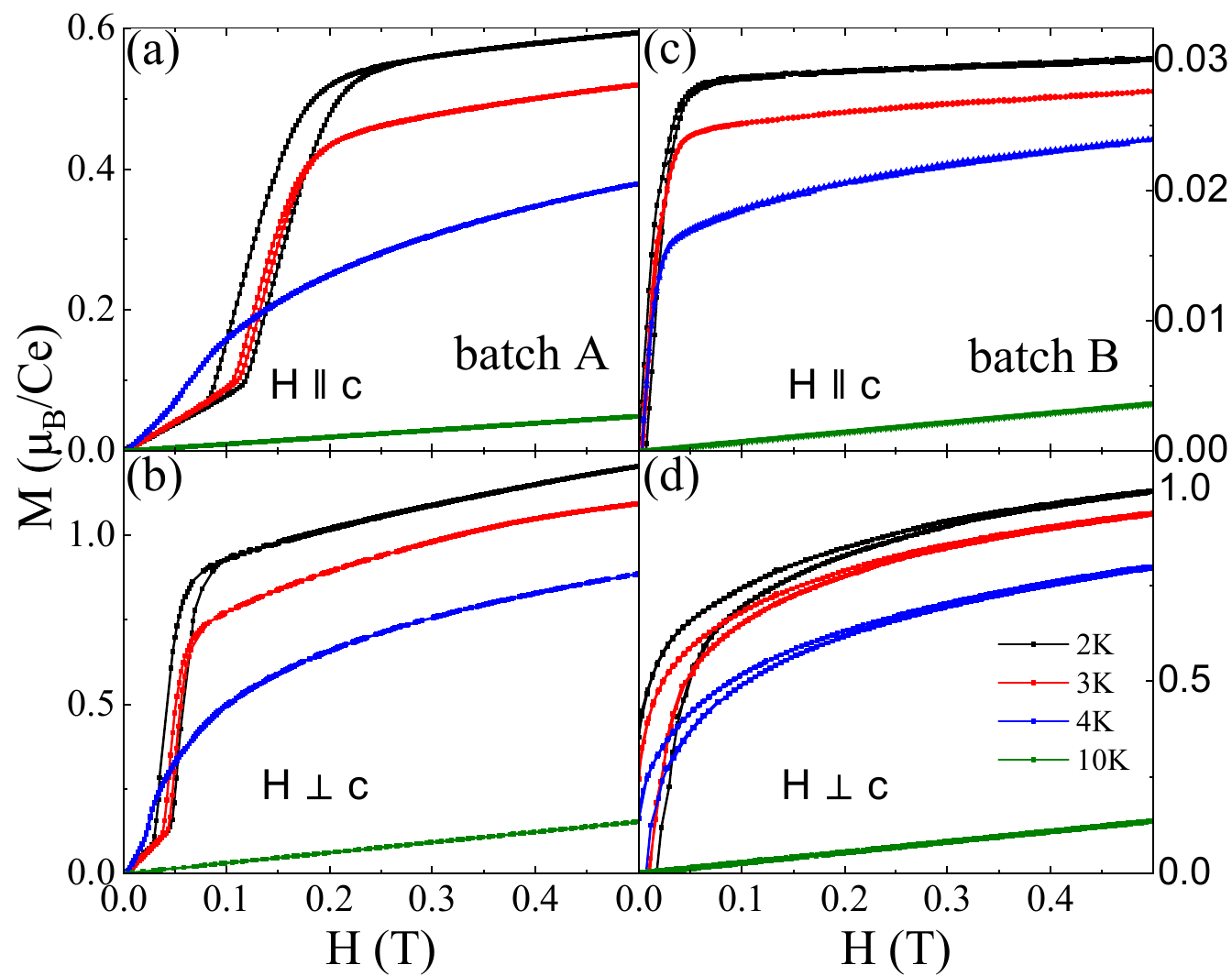}
    \captionof{figure}{Magnetization of CeZn$_{2-x}$Ga$_{2+x}$ compared between two batches. Low temperature magnetization for field (a) parallel to and (b) perpendicular to the $c$-axis-axis at fixed temperatures for batch A. Low temperature magnetization for field (c) parallel to and (d) perpendicular to the $c$-axis-axis at fixed temperatures for batch B.}
    \label{fig:mag}
\end{figure}

\subsection{Magnetic properties}

\begin{figure*}[!ht]
    \adjincludegraphics[height=7.1cm,trim={1.8cm 1cm 0.5cm 0.5cm},clip]{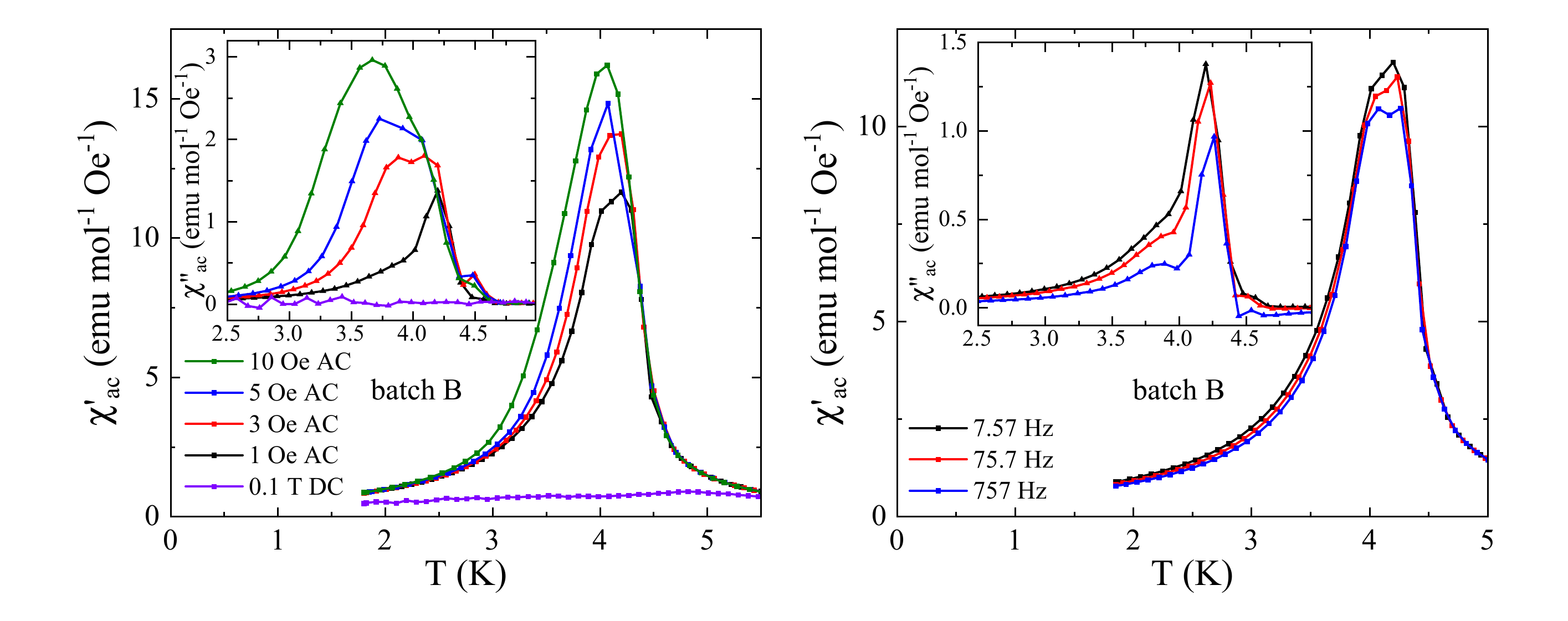}
    \captionof{figure}{AC susceptibility measurements of CeZn$_{2-x}$Ga$_{2+x}$ batch B. (a) $\chi'$ and $\chi''$ (inset) with varying fields - purple with a 0.1~T DC field and 1 Oe AC pulse. Other curves are at nearly zero DC field. All curves taken at 7.57 Hz pulses. (b) $\chi'$ and $\chi''$ (inset) as a function of frequency. All curves at nearly zero DC field and 1 Oe AC pulse.}
    \label{fig:acChi}
\end{figure*}

To study the nature of the observed transitions, we have performed detailed magnetic susceptibility, magnetization and AC susceptibility measurements. We observe the presence of a magnetic transition at low fields around 4 K for all grown batches of the material, shown in Fig.\ref{fig:chi}. In all batches, we notice a larger magnitude of the magnetic susceptibility for $H \perp c$ than in $H \parallel c$. For batch A, the transition temperature shifts faster over the low-field range for field perpendicular than parallel. For batch batch B, the transition temperature does not shift as quickly when compared to batch A, though we do still see a suppression as a function of increasing field. The differences between the two batches are exemplified further in the magnetization data shown in Fig. \ref{fig:mag}. In batch A, we observe the presence of a metamagnetic transition in M(H) signified by a ferromagnetic-like hysteresis loop that isn't centered about zero field for both field parallel and perpendicular to the $c$-axis, though the onset and closing field values differ depending on the orientation of the sample. In contrast, batch B crystals show a hysteresis loop centered about zero, consistent with a ferromagnetic transition. Although not shown in this work, we note that the magnetic properties of this material can be very sensitive to the purity of Ce used during crystal synthesis as well as to the details of the growth process, such as cooling rates and annealing times. We have reproduced the non-zero hysteresis seen in batch A in a batch annealed for several weeks with high purity Ce, suggesting that this feature is not a simple impurity effect. We note that high temperature Curie-Weiss fits to the inverse susceptibility data show that there is a competition between ferromagnetic and antiferromagnetic correlations within these samples, as $\theta \approx-34.6$ K for $H \parallel c$ and $\approx 16.1$ K for $H \perp c$ for a batch A crystal shown in SI.

The observed anisotropic magnetic properties and the slight differences in response between batches can be understood to be a result of the local disorder of Ce atomic positions within the crystalline samples. Since Ce is primarily responsible for the magnetic properties of the material, changes in its local position within the structure can affect the nature of the metamagnetic transitions and the dominant fluctuations within the sample. The bifurcation observed in the contrast between the ZFC and FC $\chi_{dc}(T)$ measurements indicates that the material is likely in a spin glass state at low temperatures. We further believe the magnetic properties of the compound are heavily affected by the shifts in the local Ce atomic positions due to the effects annealing has on the properties of the crystals. Leaving the crystals in the furnace to anneal gives time for the Ce positions to shift, allowing for a significant change in the magnetic properties. The exact positions of the Ce atoms cannot be controlled with the growth procedure, but this further supports the sensitivity of the magnetism in this system and the potential for tunability desired in magnetic devices.

In order to investigate the nature of the possible spin glass state, dynamic magnetic properties need to be studied. To do so, we performed $\chi_{ac}$ measurements, shown in Fig.\ref{fig:acChi}. We observe the presence of a peak in $\chi'$ and $\chi''$ which signifies a time-dependent response starting slightly above the magnetic transition observed in DC magnetization measurements. This peak disappears upon introduction of a background 0.1~T DC field which flips the spins to align with the field. Interestingly, we see a notable shift in the peak-center temperature as a function of the amplitude of the AC field modulation in (c). This shows how sensitive the magnetism is in this CeZn$_{2-x}$Ga$_{2+x}$ system to small perturbations, which could be of use in sensitive devices. The $\chi'$ and $\chi''$ signals as a function of field pulse frequencies are shown in (b). We observe a slight but notable change in amplitude of the $\chi'$ peak that decreases as a function of increasing frequency, but the shift is more pronounced in $\chi''$ where both the peak position and amplitude shift with increasing frequency. This frequency dependence is a direct measure of the dynamical nature of the spin glass transition.

\section{Discussion and conclusion}

The neutron diffraction results together with reverse Monte Carlo analysis suggest that there are local fluctuations of the Ce atom position in the unit cell throughout the single crystal that may lead to the magnetic properties observed. A previous report on CeZn$_2$Ga$_2$\cite{verbovytskyy_novel_2010} did not see similar magnetic responses in their samples, though other measured crystals in the paper, namely (Pr,Nd,Sm)Zn$_2$Ga$_2$, did show magnetic transitions in the susceptibility scans. Notably, our single crystals are compositionally different from the report, with some mixing of Zn and Ga atoms that results in a formula closer to 1-1-3 than 1-2-2. This mixing of Zn and Ga atoms, however, is not expected to result in such significant magnetic differences between the samples. Neither Zn nor Ga typically order magnetically in materials, and CeZnGa$_3$ and CeZn$_2$Ga$_2$ are isostructural. An alternative possible explanation of the observed magnetic properties in our crystals of CeZn$_{2-x}$Ga$_{2+x}$ is the presence of other rare earth impurities. Though we cannot exclude the presence of impurities, we do not believe this to be a significant factor as multiple batches of crystals grown with three different sources of Ce consistently result in the observed properties. All of these sources of Ce were 99.9\% and higher purity, although interestingly, we did see the closest resemblance to the expected 1-2-2 powder diffraction results when using Ce obtained from Ames lab. Additional details of differences between Ce sources can be found in the SI. Any chemical impurities would also have to be present in small enough amounts not to be detected in EDS, WDS, and XRF measurements. Considering that Verbovytskyy et al. used similar materials in their growth procedures, we do believe that the local structural distortions we observe to be dominant in the magnetic contribution to the observed anisotropic signals.

In this work, we report on the structural and magnetic properties of CeZn$_{2-x}$Ga$_{2+x}$ through measurements of magnetic susceptibility, neutron and X-ray diffraction, as well as electrical resistivity and heat capacity. We used reverse Monte Carlo analysis of the obtained neutron data to identify local structure fluctuations that cause significant refinement difficulties for both powder and X-ray diffraction measurements. We believe that the local disorder of the Ce atoms is responsible for the anisotropic magnetic properties observed, though further studies are warranted on the effects of chemical substitution and site mixing on the physical properties of CeZn$_{2-x}$Ga$_{2+x}$ and related compounds.

\section{Acknowledgments}

Research at the University of Maryland was supported by the Gordon and Betty Moore Foundation’s EPiQS Initiative through Grant No. GBMF9071, the U.S. National Science Foundation Grant No. DMR2303090, the Binational Science Foundation Grant No. 2022126, and the Maryland Quantum Materials Center. We acknowledge the support of the Maryland NanoCenter and its AIMLab.
This research used resources at the High Flux Isotope Reactor and Spallation Neutron Source, a DOE Office of Science User Facility operated by the Oak Ridge National Laboratory. The beam time was allocated to HB-2A on proposal number IPTS-33050, and POWGEN on mail-in proposal number IPTS-34975.

\bibliography{CeZnGa}

\end{document}

\title{Supplemental Material: Single crystal growth, structural and magnetic properties of CeZn$_{2-x}$Ga$_{2+x}$}

\author{Danila Sokratov}
\affiliation{Maryland Quantum Materials Center, Department of Physics, University of Maryland, College Park, MD 20742, USA}
\author{H. Cein Mandujano}
\affiliation{Department of Chemistry and Biochemistry, University of Maryland, College Park, MD 20742, USA}
\affiliation{Maryland Quantum Materials Center, Department of Physics, University of Maryland, College Park, MD 20742, USA}

\author{Ram Kumar}
\author{Jared Z. Dans}
\author{Phineas Sobel}
\author{Nicholas A. Crombie}
\affiliation{Maryland Quantum Materials Center, Department of Physics, University of Maryland, College Park, MD 20742, USA}

\author{Alicia Manj\'on-Sanz}
\affiliation{Neutron Scattering Division, Oak Ridge National Laboratory, Oak Ridge, TN 37831, USA}

\author{Danielle R. Yahne}
\affiliation{Neutron Scattering Division, Oak Ridge National Laboratory, Oak Ridge, TN 37831, USA}

\author{Philip Piccoli}
\affiliation{Department of Geological, Environmental, and Planetary Sciences, University of Maryland, College Park, MD 20742, USA}

\author{Peter Y. Zavalij}
\affiliation{Department of Chemistry and Biochemistry, University of Maryland, College Park, MD 20742, USA}

\author{Efrain E. Rodriguez}
\affiliation{Department of Chemistry and Biochemistry, University of Maryland, College Park, MD 20742, USA}
\affiliation{Maryland Quantum Materials Center, Department of Physics, University of Maryland, College Park, MD 20742, USA}
\author{Johnpierre Paglione}
\email{paglione@umd.edu}
\affiliation{Maryland Quantum Materials Center, Department of Physics, University of Maryland, College Park, MD 20742, USA}
\affiliation{Canadian Institute for Advanced Research, Toronto, Ontario M5G 1Z8, Canada}

\date{\today}

\maketitle

\section{Chemical Composition Analysis}
The chemical composition of the CeZn$_{2-x}$Ga$_{2+x}$ single crystals was verified using X-ray Fluorescence (XRF), Wavelength Dispersive Spectroscopy (WDS), and Energy Dispersive Spectroscopy (EDS). Table \ref{tab:composition} summarizes the averaged atomic percentages obtained from multiple points on the crystal surface. Chemical composition analysis is relatively consistent across batches, averaged data below is for Batch B for XRF and WDS data, and batch A for EDS. Typical error rates for these measurements were ~4-8 at\% for EDS measurements and less than 5\% for WDS and XRF data. Results for the chemical composition vary across different batches, but are typically Ce$_{x}$ZnGa$_{y}$, where x is around or slightly smaller than 1, and y is typically between 2 and 3. It is important to note that theoretical calculations predict the CeZnGa$_2$ phase is unstable. The compositional analyses are also susceptible to surface compositional differences, though we expect that to typically overshoot the Ga concentration due to the self-flux growth method of the crystals.

\begin{table}[h!]
\centering
\caption{Atomic concentration analysis of CeZn$_{2-x}$Ga$_{2+x}$ samples.}
\label{tab:composition}
\begin{tabularx}{\textwidth}{l @{\extracolsep{\fill}} c c c}
\toprule
Element & XRF (at\%) & WDS (at\%) & EDS (at\%) \\
\midrule
Ce & 19.71 & 24.40 & 21.45 \\
Zn & 25.67 & 25.25 & 29.00 \\
Ga & 54.62 & 50.35 & 47.74 \\
\bottomrule
\end{tabularx}
\end{table}

\section{Neutron Powder Diffraction}

\begin{figure}[!ht]
    \includegraphics[width=0.99\linewidth]{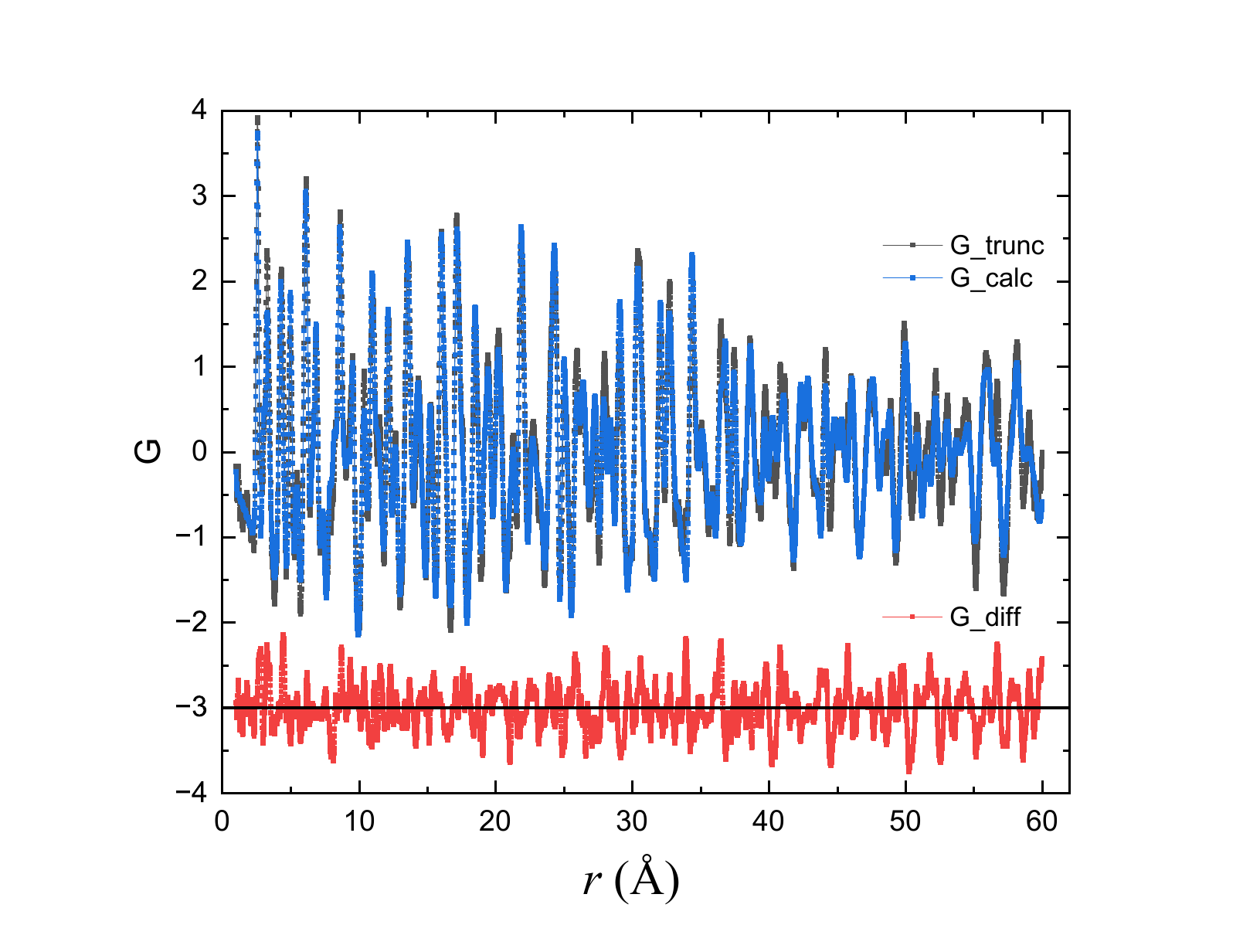}
    \captionof{figure}{Fit of CeZn$_2$Ga$_2$ refinement from Verbovytskyy et al. with NPD data from POWGEN. The difference curve has been displaced for visual clarity, with a 0-line cutting through it.}
    \label{fig:powgen-pdf}
\end{figure}

Neutron Powder Diffraction (NPD) data was collected at the POWGEN and HB-2A beamlines at ORNL. The POWGEN data with a CeZn$_2$Ga$_2$ refinement fit from Verbovytskyy et al. is shown in Fig.\ref{fig:powgen-pdf}. The difference curve shows that there are decent deviations from the averaged 1-2-2 structure, but also that it is a reasonable starting structure for the RMC calculations.

\section{X-ray Diffraction}

\begin{figure}[!ht]
    \includegraphics[width=0.85\linewidth, height=0.8\textheight, keepaspectratio]{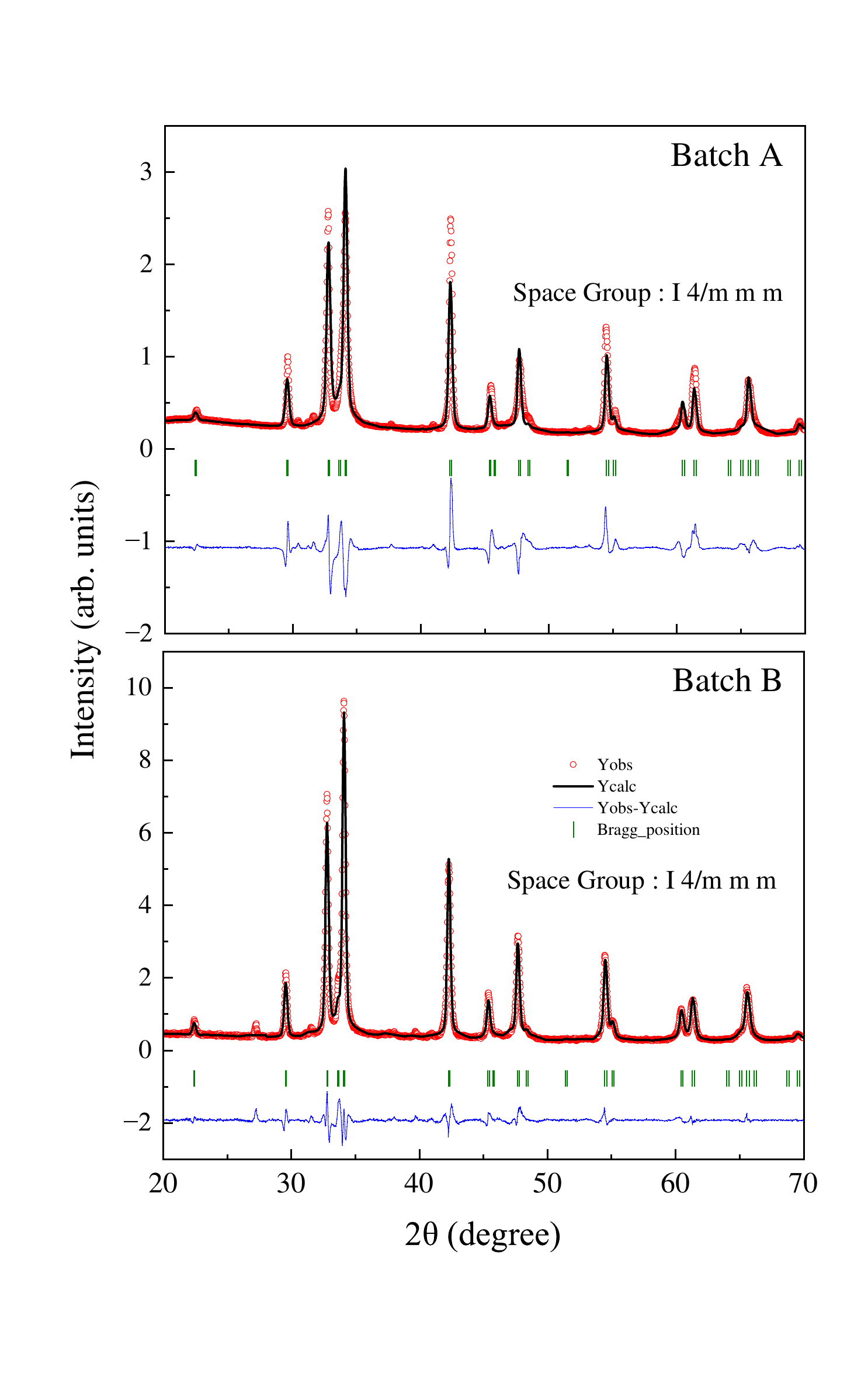}
    \captionof{figure}{Powder X-ray diffraction refinement of nominally CeZn$_2$Ga$_2$ crystals of batch A (top) and batch B (bottom). Results were fit to the CeZn$_2$Ga$_2$ phase obtained by Verbovytskyy et al.}
    \label{fig:pXRD}
\end{figure}

Single crystal refinement data and structural solution results were taken with a D8 Venture with Photon III detector. Structural solution results are presented in Table \ref{tab:crystal_structure}, while the structural parameters are shown in Table \ref{tab:atomic_params}. Powder diffraction data was taken on a Rigaku MiniFlex600 X-ray diffractometer with X-ray  diffraction (XRD) patterns taken on a powder with monochromated Cu-K$\alpha$ radiation ($\lambda \sim$ 1.5406 \AA). Comparison of diffraction patterns of two batches discussed in the main text is shown in Fig.\ref{fig:pXRD}. The structure from ref.\cite{verbovytskyy_novel_2010}, corresponding to CeZn$_2$Ga$_2$, is used to fit the datasets. The few peaks unexplained by the 1-2-2 structures can be mainly explained by the presence of Ga flux, though we note there are additional differences that could be stemming from disorder in the crystals. We note that powder X-ray diffraction patterns were taken on crystals from many other batches not discussed in the main text, which primarily see a difference in the relative intensities between the peaks corresponding to the 1-2-2 structure, signifying that there is structural variation that we attribute to the positional disorder of the Ce atom.

\begin{table}[h!]
\nopagebreak
\caption{Single-crystal XRD structural solution results for $\text{CeZn$_2$Ga$_2$}$ phase.}
\label{tab:crystal_structure}
\begin{tabularx}{\textwidth}{X l r}
\hline\hline
\addlinespace[0.5em]
Cryst system & \multicolumn{2}{r}{Tetragonal} \\
Space group & \multicolumn{2}{r}{$I4/mmm$} \\
$a$ & \multicolumn{2}{r}{4.2664(6) \AA} \\
$c$ & \multicolumn{2}{r}{10.684(2) \AA} \\
Volume & \multicolumn{2}{r}{194.47(7) \AA$^3$} \\
$Z$ & \multicolumn{2}{r}{2} \\
Calculated density & \multicolumn{2}{r}{7.007 g/cm$^3$} \\
$F(000)$ & \multicolumn{2}{r}{360.0} \\
Crystal size (mm) & \multicolumn{2}{r}{0.11 $\times$ 0.1 $\times$ 0.065} \\
Radiation & \multicolumn{2}{r}{Mo$K_{\alpha}$ ($\lambda = 0.71073$ \AA)} \\
$2\Theta$ range & \multicolumn{2}{r}{7.628$^\circ$ to 59.968$^\circ$} \\
Refl collected & \multicolumn{2}{r}{1249} \\
Independent refl & \multicolumn{2}{r}{107} \\
G.O.F. on $F^2$ & \multicolumn{2}{r}{1.196} \\
Final $R$ indexes & $R_1$ & 0.0232 \\
$[\text{I} \geqslant 2\sigma(\text{I})]$ & $wR_2$ & 0.0578 \\
Final $R$ indexes & $R_1$ & 0.0246 \\
{[all data]} & $wR_2$ & 0.0583 \\
Temperature & \multicolumn{2}{r}{298(2) K} \\
\addlinespace[0.5em]
\hline\hline
\end{tabularx}
\end{table}

\begin{table}[h!]
\centering
\caption{Structural parameters for $\text{CeZn$_2$Ga$_2$}$ obtained from single-crystal XRD results at 298 K.}
\label{tab:atomic_params}
\begin{tabular}{lccccccccccc}
\hline\hline
\addlinespace[0.5em]
Atom & Wyc. Pos. & $x/a$ & $y/b$ & $z/c$ & Occ. & $U_{11}$ & $U_{22}$ & $U_{33}$ & $U_{23}$ & $U_{13}$ & $U_{12}$ \\
\addlinespace[0.3em]
\hline
\addlinespace[0.4em]
Ce & $2b$ & 0 & 0 & 1/2 & 1 & 0.0078(4) & 0.0078(4) & 0.0088(5) & 0 & 0 & 0 \\
Zn & $4d$ & 1/2 & 0 & 1/4 & 1 & 0.0165(5) & 0.0165(5) & 0.0076(6) & 0 & 0 & 0 \\
Ga & $4e$ & 1/2 & 1/2 & 0.38451(14) & 1 & 0.0129(5) & 0.0129(5) & 0.0089(7) & 0 & 0 & 0 \\
\addlinespace[0.5em]
\hline\hline
\end{tabular}
\end{table}

\section{Single crystal growth}

The discussion in the main text focuses on two batches of CeZn$_{2-x}$Ga$_{2+x}$, but many more batches of this compound were grown to test differences in properties. The major differences between batches include elemental ratios and the sources of elements used in the growths, as well as a couple of batches where quenching or annealing were done to test their effects on properties. Each of these will be discussed in detail below. All of the batches that produced crystals were visually similar but with differences like clumping and amounts of flux stuck to the surface. Yields were approximately equivalent between batches that were successful.

\subsection{Elemental ratios}

The original growth attempt of nominal CeZn$_2$Ga$_2$ was of three batches using identical elements but different ratios to test the need for and effectiveness of fluxes. This included ratios of Ce:Zn:Ga of 1:2:2 (stoichiometric), 1:10:10 (mixed flux), and 1:2:10 (Ga flux). The stoichiometric batch had signatures of reactions, but there were still visible Zn pellets in the crucible without any crystalline formations. Both the mixed flux and Ga flux batches produced crystals that were "dirtied" with a silvery-black coating of Ga on the surface that could be scraped off, but otherwise clung to the crystals. The crystals in the 1:10:10 batch were between 1 and 4 $mm$ on each side and around 400 $\mu m$ in thickness, whereas the crystals in the 1:2:10 batch primarily clumped in two large formations around 6 $mm$ by 4 $mm$ and around 1 $mm$ thickness. The powder diffraction patterns between these two batches were qualitatively similar to ones seen in \ref{fig:pXRD}, but with similar differences in peak intensities of the major peaks near $2\theta$ $\approx$ $34\degree$ and $\approx$ $42\degree$.

Additional tested ratios include Ce:Zn:Ga 1:2:20, 1:5:5 and 1:10:2. 1:2:20 yielded crystals, but there was no noticeable improvement over other ratios in yields or X-ray powder diffraction patterns. 1:5:5 had no yields but the elements had reacted. 1:10:2 provided crystals that were trapped in a cylinder of malleable flux. Those crystals exhibited signs of a different symmetry group and are not discussed in this work.

\subsection{Sources of elements}

For all of the CeZn$_{2-x}$Ga$_{2+x}$ growths, the same source of Zn pellets was used. We tested a different source of Ga after the first few growths due to concerns of contamination after long-time storage, but found there were not significant differences with the newer bottle of Ga, besides ease of use. Major differences were in source of Ce used for the growths. All of the different Ce sources attempted in these growths were stored in the same glovebox to minimize oxidation, and were cut and weighed inside the glovebox. One source was from Alfa Aesar, listed at 99.9$\%$ metals basis; the other source was from Ames lab, and though the exact purity was not noted, Ames lab purified rare earth materials are notorious for being very high purity and, importantly, more isolated from contamination of other rare earth materials.

After many batches of testing, we found that the powder X-ray diffraction pattern deviation from the expected 1-2-2 phase in the intensity of the highest peaks was less when using Ames lab Ce compared to the Alfa Aesar source. Other complications, like presence of Ga inclusions or flux contamination in the powder, had the effect of adding other impurity peaks to the diffraction pattern, but those were largely unaffected by the Ce source. Additionally, we attempted cleaning the Ce before loading it in a growth by arc-melting the pieces in hopes of breaking down the oxidation layer, but found that it had the opposite effect, resulting in failed-yield batches.

\subsection{Annealing}

\begin{figure}[!ht]
    \includegraphics[width=0.95\linewidth, height=0.8\textheight, keepaspectratio]{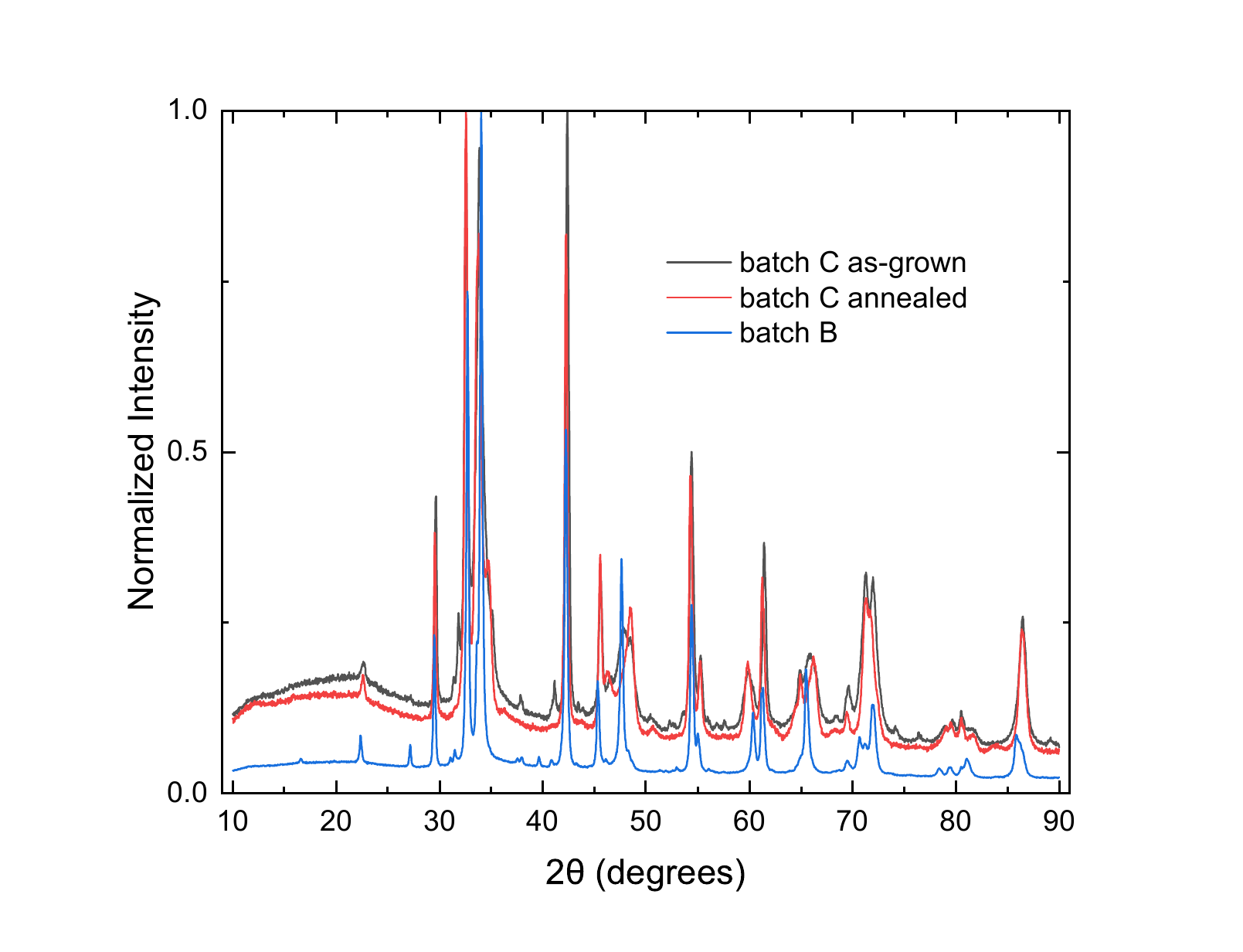}
    \captionof{figure}{Comparison of annealing effects on grown crystals of CeZn$_{2-x}$Ga$_{2+x}$. Blue curve (batch B) is one of the batches discussed in the main text and has the best qualitative fit to previous structural reports of CeZn$_2$Ga$_2$. Batch C was grown under nominally identical conditions.}
    \label{fig:anneal}
\end{figure}

One important test was whether annealing could help improve the powder X-ray diffraction patterns to be closer to the expected reported 1-2-2 phase. 4 large crystals from a batch were sealed in a quartz ampule with no transport agent, loaded into a horizontal furnace, and set at max ramping rate to go to 400$\degree C$, dwell there for 20 days, then were quenched in water to quickly cool the crystals. Visually, post-annealing crystals looked identical to pre-annealing. Interestingly, there are differences in the powder X-ray diffraction patterns between crystals from the batch pre- and post-annealing, but it does not seem like the annealing was getting the diffraction pattern closer to the expected 1-2-2 phase from ref.\cite{verbovytskyy_novel_2010}, as shown in Fig.\ref{fig:anneal}. We note that since the comparison measurement was done on two different crystals from the batch, it is unclear whether the entirety of the diffraction pattern difference can be attributed to annealing of the sample, but from previous powder diffraction measurements, we do not see significant inter-batch variation compared to differences between batches.

\bibliography{CeZnGa}